\documentclass[a4paper,12pt]{article}
\usepackage[dvips]{graphicx}
\usepackage{amsmath}
\def\BPK{B \to \phi K_S}
\def\BEK{B \to \eta' K_S}
\def\Spk{S_{\phi K}}
\def\Sek{S_{\eta' K}}
\def\pek{\phi(\eta') K_S}
\def\10{\bf 10}
\def\5b{\bf {\bar{5}}}
\def\bra{\langle}
\def\ket{\rangle}
\def\gsim  {\hspace{0.3em}\raisebox{0.4ex}{$>$}\hspace{-0.75em}\raisebox{-.7ex}{$\sim$}\hspace{0.3em}}
\def\lsim  {\hspace{0.3em}\raisebox{0.4ex}{$<$}\hspace{-0.75em}\raisebox{-.7ex}{$\sim$}\hspace{0.3em}}

\begin{document}


\vspace{4ex}

\begin{center}
{\large \bf
CP asymmetries of $B \to \phi K_S$ and $B \to \eta' K_S$ \\in SUSY GUT Model with Non-universal Sfermion Masses
}

\vspace{6ex}

\renewcommand{\thefootnote}{\alph{footnote}}
S.-G. Kim\footnote{e-mail: sunggi@eken.phys.nagoya-u.ac.jp},
N. Maekawa\footnote{e-mail: maekawa@eken.phys.nagoya-u.ac.jp},
A. Matsuzaki\footnote{e-mail: akihiro@eken.phys.nagoya-u.ac.jp}, 
\\K. Sakurai\footnote{e-mail: sakurai@eken.phys.nagoya-u.ac.jp},
and 
T. Yoshikawa\footnote{e-mail: tadashi@eken.phys.nagoya-u.ac.jp}

\vspace{4ex}
{\it Department of Physics, Nagoya University, Nagoya 464-8602, Japan}\\

\end{center}

\renewcommand{\thefootnote}{\arabic{footnote}}
\setcounter{footnote}{0}
\vspace{8ex}


\begin{abstract}

We analyze $CP$ asymmetries of $B \to \phi K_S$ and $B \to \eta' K_S$
in a supersymmetric grand unified theory in which only the third
generation sfermions contained in ${\bf 10}{\rm (Q, U^c, E^c)}$ 
of $SU(5)$ can have a different mass from the others. 
One of the advantages of this nonuniversal mass model is that the
first two generation sfermion masses can be large whereas both 
(left and right handed) stops are light so as to stabilize the weak scale.
Therefore, we studied a minimal supersymmetric standard model
parameter region in which a fine tuning in Higgs sector is relaxed 
owing to light masses of stops, gluino and higgsinos.
In such a parameter region, the chargino contribution is as important
as the gluino one.
We show that the $CP$ asymmetries of $B \to \phi K_S$ and $B \to \eta'
K_S$ can deviate from their standard model predicted values 
by ${\cal O}(0.1)$ because of constructive interference between gluino and chargino contributions.
\end{abstract}

\vspace{2cm}


\section{Introduction}

Supersymmetry (SUSY) is one of the most promising candidates for physics beyond the Standard Model (SM).
The minimal supersymmetric standard model (MSSM) not only provides a
solution to the gauge hierarchy problem in the SM but also has some
attractive features, for example, the gauge coupling unification, 
the radiative electroweak symmetry breaking, and giving dark matter candidates as the lightest superparticle.   

However, the MSSM has serious problems in SUSY breaking sector.
If we introduce generic SUSY breaking terms, they induce very large
amplitudes for Flavor Changing Neutral Current (FCNC) and $CP$
violating processes to satisfy the various experimental 
constraints\cite{Ellis:1981ts-Barbieri:1981gn, Hagelin:1992tc, Gabbiani:1996hi}.
To avoid the problems the universal soft sfermion masses are often assumed at some scale\cite{Nilles:1983ge, Kane:1993td}.

The assumption of the universal soft masses is not necessarily required.
For example, since the FCNC constraints for the first two generation
field are much severer than for the third generation fields, the
universal soft masses sometimes have been imposed only for the first
two generation sfermion, which are realized if we impose non-Abelian
horizontal symmetry, for example, $U(2)$, under which the first two 
generation fields are doublets and the third generation fields are singlets \cite{horizontal}.
However, if the diagonalizing matrix of the fermion has 
Maki-Nakagawa-Sakata \cite{MNS} (MNS)-{\it like} large mixings, such
sfermion mass spectrum leads to very large FCNC to be consistent 
with the experimental bound. Therefore, such nonuniversality should 
be introduced only for the sfermion whose fermionic superpartners have 
the diagnalizing matrix with the Cabibbo-Kobayashi-Maskawa \cite{CKM} (CKM)-{\it like} small mixings.

In $SU(5)$ SUSY Grand Unified Theory (GUT), it is a reasonable
assumption that the diagonalizing matrices of {\bf 10} 
have the CKM-{\it like} small mixings, while those of ${\bf \bar{5}}$ 
have the MNS-{\it like} large mixing, because the {\bf 10} includes 
the doublet quarks, and ${\bf \bar{5}}$ includes the doublet leptons.
Under this assumption, only the sfermions included in the third 
generation of the {\bf 10} (${\bf 10}_3$), can have different masses 
from the others without conflicting with experimental constraints from various FCNC processes.

We summarize several characteristic features of the nonuniversal sfermion masses.
First of all, the rotation matrices for {\bf 10}, which make fermions
to mass eigenstates are expected to have the CKM-{\it like} small
mixings, since {\bf 10} involves quark doublet.
Therefore, large off-diagonal entries for the {\bf 10} sfermion 
mass matrices do not arise after this rotation, even if the initial soft masses are not universal.
Moreover, FCNC constraints among 1-3 or 2-3 generations are not so
severe compared with those of 1-2 generations.
Therefore, we can expect that nonuniversality for the third generation
does not conflict with the FCNC constraints.
Moreover, it is important that we can expect larger FCNC for the third generation fields than in the usual universal sfermion mass case
as discussed in Ref.\cite{E6LFV}.
Second, the naturalness of the Higgs mass in MSSM requires that the
gaugino masses, the higgsino mass, and the stop masses must be around the weak scale.
In the nonuniversality case, we can take the larger masses for all the
sfermions except for the sfermions of the ${\bf 10}_3$ without
conflicting with the naturalness arguments, because both (left and right handed) stops are included in the ${\bf 10}_3$.
In principle, we can take such mass larger than 1 TeV, which can relax the various constraints from FCNC and electric dipole moments (EDM), etc.
Again, the FCNC related with the third generation fields can become relatively large, which may be detectable in future experiments.

Finally, such a sfermion mass spectrum can be realized in the anomalous $U(1)$ SUSY-breaking models\cite{Nakano:1994sw},
flavor-mediated SUSY breaking models\cite{Kaplan:1998jk}, 
radiatively driven models with specific boundary conditions\cite{Feng:1998iq},
extra dimension models,\cite{Haba:2002vd}
and models with discrete symmetry,\cite{Hamaguchi:2002vi}
and is naturally derived from ${\rm E}_6$ SUSY GUT model \cite{E6, E6r} with $SU(2)$ or $SU(3)$ horizontal symmetry \cite{E6Hori}.


It is expected that FCNC processes among 2-3 generations,
particularly, the $b_L \to s_L$ or $\tau_R \to \mu_R$  transition rate
become relatively large in the nonuniversal sfermion mass model.
In the previous paper \cite{E6LFV}, we studied $\tau \to \mu \gamma$ 
and $\mu \to e \gamma$ processes and showed that these lepton flavor 
violating processes can have very large branching ratios and that these processes may be discovered in near future experiments. 
However, in the quark sector, this model gives similar predictions for the
various FCNC processes to those in the model with the universal sfermion masses because
the off-diagonal elements of the squark mass matrix obtained from the nonuniversal sfermion masses are of the same order 
as those obtained by the renormalization group equations (RGEs) from
the universal sfermion masses. 
The difference between these models appears in the $CP$ violating 
processes because, in the nonuniversal sfermion mass model, there are new $CP$ phases except
for the Kobayashi-Maskawa (KM) phase. Therefore, it is interesting to
study the $CP$ violating processes in the nonuniversal sfermion mass model to distinguish these models.

In this paper, we focus on the 2-3 transition in the quark sector 
and analyze the $CP$ asymmetries of $B \to \phi K_S$ and $B \to \eta' K_S$ \cite{BphiK}.
These observables attracted attention a few years ago since the Belle
Collaboration \cite{ano} reported the large deviation from 
a SM expectation $\sin 2\phi_1 (\sin 2\beta)=0.681 \pm 0.025$ \cite{HFAG}.
Their current experimental world averages are 
$S_{\phi K}=0.39 \pm 0.17$, $S_{\eta' K}=0.61 \pm 0.07$
\cite{HFAG,BPKexp} and they are almost consistent with the 
SM expectation, but it is still important to measure these observables
accurately in the search for new physics beyond the SM.
Particularly, the $CP$ asymmetries of $B \to \phi K_S$ and $B \to \eta'
K_S$ are interesting for the model with the nonuniversal masses because of the following reasons.
i) Since SUSY contribution to $CP$ violation in $b_L \to s_L$
transition can be large in this model while the SM has little 
$CP$ phase in the $b \to s$ processes, large deviations of $CP$ asymmetries from the SM are expected.
ii) Both experimental central values of $B \to \phi K_S$ and $B \to \eta' K_S$ negatively deviate from the SM expectation.
It is known that such deviations favor the models with only LL
mixing (not RR mixing) in down-type squark sector \cite{EMYK}.
This gives a strong motivation to analyze the $CP$ asymmetries 
of $B \to \phi K_S$ and $B \to \eta' K_S$ in this nonuniversal 
sfermion mass model, because the ${\bf 10}$ contain left-handed down quarks and not right-handed down quarks.
iii) As in the processes of $\tau \to \mu \gamma$ and 
$\mu \to e\gamma$, the deviations of $CP$ asymmetries of 
$B \to \phi K_S$ and $B \to \eta' K_S$ from the SM expectation strongly depend on the sfermion masses of the ${\bf 10}_3$.
Therefore, if the sfermion masses of the ${\bf 10}_3$ 
are measured in the Large Hadron Collider (LHC) experiment, 
the predictions of the branching ratios of the LFV processes 
and the $CP$ asymmetries can be more definite.
Even if the sfermion masses of the ${\bf 10}_3$ are not measured 
by LHC, we can check the consistency of this model by comparing with the LFV processes and the $CP$ asymmetries.

This paper is organized as follows.
In the next section, we will specify the model and discuss the flavor and $CP$ violating couplings in this model.
After giving definitions and notations to analyze $CP$ asymmetries in \S\,3, the results of numerical study are given in \S\,4.
In \S\,5 we will discuss some of the constraints of this model.
We summarize this paper in \S\,6.

\section{$CP$ violation in the nonuniversal sfermion mass model}

In this section, we discuss flavor and 
$CP$ violating couplings related with $CP$ asymmetries of $b \to s$ processes in the nonuniversal mass model.
First, we assume that the SM gauge group ${SU(3)_C \times SU(2)_L
\times U(1)_Y}$ is unified into $SU(5)$ 
and the mediation of SUSY breaking respects the $SU(5)$ gauge symmetry.
\footnote{Note that the following arguments can be applied to unified models with larger unified gauge groups, for instance, $SO(10)$ or ${E_6}$.}
We start from the following Lagrangian at GUT scale, 
\begin{eqnarray}
{\cal L}&=&u^c_{Ri} Y^u_{ij} q_j H_u + d^c_{Ri} Y^{d,e}_{ij} q_j H_d +\cdots \nonumber
\\&+& \tilde{u}_{Ri}^* (m_{\bf 10}^2)_{ii} \tilde{u}_{Ri} + \tilde{q}_i^* (m_{\bf 10}^2)_{ii} \tilde{q}_i+\tilde{d}_{Ri}^* (m^2_{\bf \bar{5}})_{ii} \tilde{d}_{Ri} +\cdots,
\end{eqnarray}
where $q_i$, $u^c_{Ri}$, $d^c_{Ri}$ ($i=1,2,3$) are quark doublets, right-handed up quarks, and right-handed down quarks, respectively, and the fields with tildes denote sfermions. 
As mentioned in the Introduction, we assume the following forms for the sfermion mass matrices.
\begin{equation}
m_{\bf 10}^2 = \begin{pmatrix}
m_0^2&&\\
&m_0^2&\\
&&m^2_{30}
\end{pmatrix},
~~~~~~~~~~~~
m^2_{\bf \bar{5}} = \begin{pmatrix}
m^2_0 &&\\
&m^2_0 &\\
&&m^2_0
\end{pmatrix},
\label{special mass}
\end{equation}
By redefining the superfields, we can diagonalize the up type Yukawa matrix.
\begin{equation}
U_R \to V_{U} U_R,~~~~D_R \to V_{D} D_R,~~~~Q \to V_Q Q.
\end{equation}
\begin{eqnarray}
{\cal L}&=&u^c_{Ri} (\Hat{Y}^u_0)_{ii} u_{Li} H_u^0 + d^c_{Ri} (\Hat{Y}^d_0 V^{CKM\dagger}_0)_{ij} d_{Lj} H_d^0 +\cdots \nonumber
\\&+& \tilde{u}_{Ri}^* (m_{\tilde{u}_R0}^2)_{ij} \tilde{u}_{Rj} +\tilde{q}_i^* (m_{\tilde{q}0}^2)_{ij} \tilde{q}_j
+\tilde{d}_{Ri}^* (m^2_{\bf \bar{5}})_{ii} \tilde{d}_{Ri} +\cdots,
\label{LGUT}
\end{eqnarray}
where $\hat{Y}^f_0$ and $V^{CKM}_0$ are the real and diagonal Yukawa matrices and CKM matrix at GUT scale, respectively.
$V_Q$ is a unitary matrix of quark doublets and it is roughly expected as
\begin{equation}
V_{Q} \sim \begin{pmatrix}
1&\lambda &\lambda^3\\
\lambda &1&\lambda^2\\
\lambda^3 &\lambda^2 &1
\end{pmatrix},
\label{VQ}
\end{equation}
where $\lambda$ is Cabibbo angle $(\lambda =0.22)$ and we omit the ${\cal O}(1)$ coefficients and phases.
Then, $m_{\tilde{q}0}^2$ is estimated as  
\begin{eqnarray*}
m_{\tilde{q}0}^2 = V^{\dagger}_Q m^2_{\bf 10} V_Q 
\sim
\begin{pmatrix}
m_0^2&&\\
&m_0^2&\\
&&m_{30}^2 \end{pmatrix}+m^2_{FC0},
\end{eqnarray*}
\begin{eqnarray}
m^2_{FC0} &=& V_Q^{\dagger} \begin{pmatrix}
0 & & \\
 & 0& \\
 & &m_{30}^2-m_0^2
\end{pmatrix} V_Q
- \begin{pmatrix}
0&&\\
&0&\\
&&m_{30}^2-m_0^2
\end{pmatrix}
\nonumber \\ &\sim&
(m_{30}^2-m_0^2)\begin{pmatrix}
\lambda^6&\lambda^5 &\lambda^3\\
\lambda^5 &\lambda^4&\lambda^2\\
\lambda^3 &\lambda^2 &0
\end{pmatrix}.
\label{mFC}
\end{eqnarray}
Here, we use $m^2_{\bf 10} = m^2_0 {\bf 1} + {\rm diag}(0,0,m^2_{30}-m^2_0)$ and the unitarity properties of $V_Q$.
Since we have already used the freedom of phase rotation of quark fields to reduce the number of phases in $V^{CKM}_0$ into one,
the coefficients in $V_Q$ have ${\cal O}(1)$ phases generically.
Therefore, the off-diagonal entries of $m^2_{\tilde{q}0}$ and $m_{FC0}^2$ can have new ${\cal O}(1)$ phases.
  
To calculate $CP$ asymmetries of $B \to \phi(\eta') K_S$, 
we need the low energy sfermion mass matrices $m^2_{\tilde{q}}$ and $m^2_{\tilde{d}_R}$, which are obtained by renormalization group equation (RGE).
However, off-diagonal entries of sfermion mass matrices do not change significantly in RGEs between GUT and weak scale, for example, 
\begin{eqnarray}
(m_{\tilde{q}}^2)_{32}(\mu_{weak}) &\sim& \exp\bigl[ \frac{|Y_{33}^u|^2}{2} \frac{1}{(4 \pi)^2} \log(\mu_{weak}^2/\Lambda^2_{GUT}) \bigr] \times (m_{\tilde{q}}^2)_{32}(\Lambda_{GUT}) \nonumber \\
&\sim& 0.8 \times (m_{\tilde{q}}^2)_{32}(\Lambda_{GUT})\,.
\label{rgemq}
\end{eqnarray}
In our analysis, we do not fix the off-diagonal entries at GUT scale
(We can estimate only their order.), so we absorb the high and 
low energy differences of the $(m_{\tilde{q}}^2)_{32}$ in ambiguity of $(m_{\tilde{q}}^2)_{32}(\Lambda_{GUT})$.
For other relevant off-diagonal entries, the changes between GUT 
and weak scale are rather small owing to small Yukawa couplings.

Therefore, we calculate the diagonal components of these mass matrices by RGEs.
Finally, we go to the super CKM basis by rotating the superfields as $D_L \to V_{CKM} D_L$ at low energy. 
Then, low energy sfermion mass matrices are obtained as 
\begin{eqnarray}
m_{\tilde{u}_L}^2 &\sim& \begin{pmatrix}
m^2&&\\
&m^2&\\
&&m_3^2
\end{pmatrix}
+m^2_{FC},
\label{mu}
\\
m_{\tilde{d}_L}^2 &\sim& V_{CKM}^{\dagger} 
\begin{pmatrix}
m^2&&\\
&m^2&\\
&&m_3^2
\end{pmatrix} V_{CKM}
+V_{CKM}^{\dagger} m^2_{FC} V_{CKM} ,
\label{md}
\end{eqnarray}
$m_{\tilde{d}_R}^2$ does not have off-diagonal entries.
Note that since a component $(V_{CKM})_{32}$ does not have a large
$CP$ phase, the first term in Eq.(\ref{md}) does not contribute to the parameter ${\rm Im}[(m_{\tilde{d}_L}^2)_{32}]$.
Thus, the following approximate relation is obtained.
\begin{equation}
{\rm Im}[(m_{\tilde{d}_L}^2)_{32}] \simeq {\rm Im}[(m_{\tilde{u}_L}^2)_{32}] \sim (m^2_{30}-m^2_0)\lambda^2 \sin \theta_{SUSY},
\label{cpvs}
\end{equation}
where $\theta_{SUSY}$ is a phase of $(m_{FC})_{32}$.
This relation means that if the gluino contribution is maximized by the SUSY phase, then the chargino contribution is also maximized.
Thus, strong interference between gluino and chargino contributions is expected in this model.

\section{$CP$ asymmetry of $B \to \phi K_S$ and $B \to \eta' K_S$}
In this section, we give a review of the well known method 
of effective field theory formalism to calculate the $CP$ asymmetries of $B \to \phi(\eta') K_S$ \cite{Buras, Khalil}.
Their time dependent $CP$ asymmetries are defined as
\begin{eqnarray}
a_{\pek}(t) &=& \frac{\Gamma(\bar{B}(t) \to \pek)-\Gamma(B(t) \to \pek)}{\Gamma(\bar{B}(t) \to \pek)+\Gamma(B(t) \to \pek)} \nonumber
\\
&=&A^{dir}_{\pek} \cos \Delta M_{B_d}t + S_{\pek} \sin \Delta M_{B_d}t,
\end{eqnarray}
where $A^{dir}_{\pek}$ and $S_{\pek}$ represent the direct and mixing $CP$ asymmetries, respectively, and their expressions are given as
\begin{equation}
A^{dir}_{\pek} = \frac{|\bar{\rho}(\pek)|^2-1}{|\bar{\rho}(\pek)|^2+1},~~~~~~~~
S_{\pek} = -\frac{2 {\rm Im} [e^{-2i\phi_1}\bar{\rho}(\pek)]}{|\bar{\rho}(\pek)|^2+1},
\label{CS}
\end{equation}
where $\phi_1$(or $\beta$) is the standard angle of the unitarity triangle, and parameter $\bar{\rho}(\pek)$ is defined as
\begin{equation}
\bar{\rho}(\pek) = \frac{\bar{A}(\pek)}{A(\pek)}.
\end{equation}
Here, $A(\pek)$ and $\bar{A}(\pek)$ are decay amplitudes of $B \to
\pek$ and $\bar{B} \to \pek$, respectively, 
which can be calculated using the effective Hamiltonian of $\Delta B=1$ transition at the low energy $\mu_b \simeq m_b$
\begin{equation}
\bar{A}(\pek)=\bra \pek |H^{\Delta B=1}_{eff} |\bar{B} \ket,~~~~~~
A(\pek)=\bra \pek |(H^{\Delta B=1}_{eff})^{\dagger} |B \ket.
\end{equation}
Effective Hamiltonian $H^{\Delta B=1}_{eff}$ is expressed using the operator product expansion(OPE) as \cite{Buras}
\begin{eqnarray}
H^{\Delta B=1}_{eff} = \sum_{p=u,c} \sum_{i=3}^{6,7\gamma,8g} \frac{G_F}{\sqrt{2}} \bigl[  \lambda_p (C_1 Q_1^{(p)} + C_2 Q_2^{(p)}) 
-   \lambda_t C_i Q_i  \bigr] + \{ Q \to \tilde{Q},\, C \to \tilde{C}
\}, \nonumber \\
\end{eqnarray}
where $\lambda_q = V_{qb} V_{qs}^*$, $C_i \equiv C_i(\mu_b)$ are the low energy Wilson coefficients.
The low energy renormalized operators $Q_i \equiv Q_i(\mu_b)$ are expressed as 
\begin{eqnarray}
&&Q_1^{(p)}=(\bar{p}_{\alpha} \gamma_{\mu} 2P_L b_{\beta})(\bar{s}_{\beta} \gamma^{\mu} 2P_L p_{\alpha}),~~~
Q_2^{(p)}=(\bar{p} \gamma_{\mu} 2P_L b)(\bar{s} \gamma^{\mu} 2P_L p), \nonumber
\\
&&Q_3=(\bar{s} \gamma_{\mu} 2P_L b) \sum_q (\bar{q} \gamma^{\mu} 2P_L q),~~~~~~
Q_4=(\bar{s}_{\alpha} \gamma_{\mu} 2P_L b_{\beta}) \sum_q (\bar{q}_{\beta} \gamma^{\mu} 2P_L q_{\alpha}), \nonumber
\\
&&Q_5=(\bar{s} \gamma_{\mu} 2P_L b) \sum_q (\bar{q} \gamma^{\mu} 2P_R q),~~~~~~
Q_6=(\bar{s}_{\alpha} \gamma_{\mu} 2P_L b_{\beta}) \sum_q (\bar{q}_{\beta} \gamma^{\mu} 2P_R q_{\alpha}), \nonumber
\\
&&Q_{7\gamma}=\frac{e}{8\pi^2}m_b \bar{s} \sigma^{\mu\nu} 2P_R F_{\mu\nu} b,~~~~~~~~~~~
Q_{8g}=\frac{g_s}{8\pi^2}m_b \bar{s}_{\alpha} \sigma^{\mu\nu} 2P_R G^{A}_{\mu\nu} T^A_{\alpha\beta} b_{\beta},
\label{Q}
\end{eqnarray}
where $\alpha$ and $\beta$ are color indices, $T^A_{\alpha\beta}$ is
$SU(3)_C$ generator, $\sigma^{\mu\nu}={i \over
2}[\gamma^{\mu},\gamma^{\nu}]$, $P_{L,R}=(1 \mp \gamma_5)/2$ 
represent the projection operators, and $\tilde{Q}_i$ is obtained from $Q_i$ by exchanging $L \leftrightarrow R$.
Here, we ignored the electroweak penguin operators $Q_{7-10}$ 
and the contributions to dipole operators which are proportional to the strange quark mass.
The matrix elements $\bra B(\bar{B}) |Q_i|\pek \ket$ are provided in Appendix A.
The low energy Wilson coefficients $C_i$ can be obtained 
from the high energy coefficients $C_i(\mu_W)$ ($\mu_W \simeq m_W$) by solving the renormalization group equations for QCD in the SM.
The solution is expressed as\cite{Buras}
\begin{equation}
C_i(\mu_b) = \sum_{j} \hat{U}_{ij}(\mu_b,\mu_W) C_j(\mu_W).
\end{equation}
Here, $\hat{U}_{ij}$ is the evolution matrix and we included the leading order in QCD. Their expressions are provided in Appendix B.
For simplicity, the matching scale is chosen as the $m_W$, but we calculate the Wilson coefficients in full theory, which includes the superparticles.
We construct each coefficient using the SM contributions
$C^{SM}_i(\mu_W)$, Charged Higgs contributions $C^{H}_i(\mu_W)$,
Gluino contributions $C^{g}_i(\mu_W)$, and Chargino contributions $C^{\chi}_i(\mu_W)$ as follows:
\begin{eqnarray}
&&C_1=C_1^{SM},~~~~~~~~~~~~~~~~~~~~~~~~~~~~~~~C_2=C_2^{SM}, \nonumber \\
&&C_3=C_3^{SM}+C_3^{g},~~~~~~~~~~~~~~~~~~~~~~~~C_4=C_4^{SM}+C_4^{g}, \nonumber \\
&&C_5=C_5^{SM}+C_5^{g},~~~~~~~~~~~~~~~~~~~~~~~~C_6=C_6^{SM}+C_6^{g}, \nonumber \\
&&C_{7\gamma} = C_{7\gamma}^{SM}+C_{7\gamma}^{H}+C_{7\gamma}^{g}+C_{7\gamma}^{\chi},~~~~~
C_{8g} = C_{8g}^{SM}+C_{8g}^{H}+C_{8g}^{g}+C_{8g}^{\chi}.
\label{Cs}
\end{eqnarray}
Here, we ignored the contributions of $C^H_3(\mu_W)$-$C^H_6(\mu_W)$ and $C^{\chi}_3(\mu_W)$-$C^{\chi}_6(\mu_W)$.
\footnote{We checked numerically that their contributions cannot be
large in our parameter region of interest.}
Since there are no flavor and $CP$ violation in right-handed down-type
squark sector in the nonuniversal mass model as mentioned in the previous section, the coefficient $\tilde{C}^g$ does not emerge.
The $\tilde{C}^H$ and $\tilde{C}^{\chi}$ can also be neglected 
because the $\tilde{C}_{7\gamma,\, 8g}$ is suppressed by the factor $m_s/m_b$ compared with $C_{7\gamma,\, 8g}$.
The difference in the final state parity of $\BPK$ and $\BEK$ leads to
the following relation.
\begin{equation}
\bra \phi \bar{K}^0 | \tilde{Q}_i | \bar{B}^0 \ket = \bra \phi \bar{K}^0 | Q_i | \bar{B}^0 \ket,
~~~~~~~
\bra \eta' \bar{K}^0 | \tilde{Q}_i | \bar{B}^0 \ket = -\bra \eta' \bar{K}^0 | Q_i | \bar{B}^0 \ket.
\end{equation}
The 1-loop order expression for the Wilson coefficients in Eq.(\ref{Cs}) are provided in Appendix B. 

Let us discuss the deviations of $\Spk$ and $\Sek$ from the SM expectation.
First, we extract the strong phase from the amplitudes as $A=e^{i\delta_s} A^{CP}$.
Since the dominant contribution to the $A(\pek)$, which is provided
from the SM is almost real, we can expand $\rho(\pek)$ 
and ${\bar \rho}(\pek)$ with $R_{\pek} \equiv {\rm Im}A^{CP}(\pek)/{\rm Re}A^{CP}(\pek)$ as follows.
\begin{equation}
\bar{\rho} = \frac{{\rm Re} A^{CP} - i{\rm Im}A^{CP}}{{\rm Re}A^{CP} + i{\rm Im}A^{CP}} \simeq 1-2iR 
- 2 R^2.
\end{equation}
Then using Eq.(\ref{CS}), we can derive
\begin{eqnarray}
S_{\pek} &=& \sin2\phi_1 + \Delta S_{\pek},
\\
\Delta S_{\pek} &=& 2[ R_{\pek} \cos2\phi_1 -  R_{\pek}^2 \sin2\phi_1],
\end{eqnarray}
where $\sin2\phi_1=0.68$, $\cos2\phi_1=-0.73$, and $\Delta S_{\pek}$ represent a deviation from the SM expectation.
The first term provides the leading contribution to $\Delta S_{\pek}$, and second term always provides the negative contribution.
Therefore, the maximum value of $|\Delta S_{\pek}|$ is given as positive $R_{\pek}$.

\section{Numerical analysis}
In this section, we show the numerical results of the deviation of $CP$ asymmetries of $\BPK$ and $\BEK$ from the SM expectation.
We are interested in the parameter region in which a tuning in the MSSM Higgs sector is relaxed owing to the light masses of stops, gluino and higgsinos, because one of the advantages of the nonuniversal mass model is that the light stops can be realized while the first two generation sfermions are heavy.  
Therefore, we basically use the following parameter set through our analysis.
\begin{eqnarray}
&&{\rm At~GUT~scale\,:} \nonumber \\
&&~~~m_{30}=100{\rm GeV},~~~m_0=1000~{\rm GeV},~~~m_{1/2}=200~{\rm GeV},~~~A^u_0=A^d_0=0~{\rm GeV}; \nonumber \\
&&{\rm at~ weak~ scale\,:} \nonumber \\
&&~~~\mu=250~{\rm GeV},~~~\tan\beta=10,~~~\theta_{SUSY}=-\pi/2,
\label{pset}
\end{eqnarray}
where $m_{1/2}$ is the universal gaugino mass and $\mu$ is the higgsino mass.
We took them as relatively small to relax the fine tuning in the MSSM Higgs sector.

At first glance, we may think that the above parameter region
conflicts with the LEP constraint for the higgs boson mass.
However, the well-known LEP constraint on the SM Higgs boson mass,
114.4 GeV cannot be applied to the ``MSSM'' lightest higgs boson 
if the MSSM lightest higgs does not have relevant coupling to the Z boson \cite{LHPlightH,lightH,KaneDrees}.
In this situation, the large quantum correction to the SM-{\it like} Higgs boson from the large stop and gluino masses is not required.
We checked that the above parameter region is numerically consistent with the LEP constraint in the literature \cite{LHPlightH}.
In the above parameter region, since charginos and stops are relatively light, it is expected that not only gluino diagram but also chargino diagrams give relatively large contribution to the $CP$ asymmetries.

In such a small $ZZh$ coupling region, it is known that all MSSM Higgs
bosons have relatively small masses\cite{LHPlightH,lightH}. 
Then, negative $\mu$ and large $\tan\beta$ are disfavored 
by the constraints from $b \to s \gamma$, $t \to H^+ b$, $B_u^+ \to \tau^+ \nu_{\tau}$, and $B_s \to \mu^+ \mu^-$ processes.
There are large contributions to the $b \to s \gamma$ 
from the charged Higgs-top loop diagram and the chargino-stop loop diagram in our parameter region.
The typical orders of magnitudes of their amplitudes are of the same
order as the SM one and the cancelation between them is therefore required.
This cancelation is realized only in the positive $\mu$ case.
At the same time, $\tan \beta$ cannot be arbitrarily large 
otherwise the top quark significantly decays into charged Higgs boson and $b$ quark ($\tan\beta \lsim 40$ is mandatory\cite{Abulencia:2005jd}
).
%
%
Moreover, the contributions from the light Higgs boson exchange
to the $BR(B_u^+ \to \tau^+ \nu_{\tau})$ and the $BR(B_s \to \mu^+ \mu^-)$ are highly sensitive to large $\tan\beta$, 
because they are proportional to $(\tan \beta)^4$ and $(\tan \beta)^6$, respectively.\cite{Hou:1992sy,Babu:1999hn}
Thus, the constraints from these processes disfavor the region in which $\tan\beta \gsim 15$\cite{Isidori:2006pk}.
From the above reasons, we do not consider the negative $\mu$ or large $\tan\beta$ region in this paper.

Before presenting the numerical results, we give a rough argument to show that we can discuss the magnitude of each contribution to $\Delta S$ separately. 
To separate each contribution, we use the following notation:
\begin{equation}
A^{CP}(\pek)=A^{SM}_{\pek}+A^{H}_{\pek}+A^{\chi}_{\pek}+A^{g36}_{\pek}+A^{g8}_{\pek},
\end{equation}
where  $A^{SM}_{\pek}$ is the SM amplitude and the remainders are
\begin{eqnarray}
&&A^{H}_{\pek}= -\frac{G_F}{\sqrt{2}} \lambda_t  [0.727 C^H_{8g}(\mu_W)]  \bra \pek |Q_{8g}| \bar{B}^0 \ket,
\nonumber \\
&&A^{\chi}_{\pek}= -\frac{G_F}{\sqrt{2}} \lambda_t  [0.727 C^{\chi}_{8g}(\mu_W)]  \bra \pek |Q_{8g}| \bar{B}^0 \ket,
\nonumber \\
&&A^{g36}_{\pek}= -\frac{G_F}{\sqrt{2}} \lambda_t \sum_{i=1}^6 \bigl[\sum_{j=1}^6 \hat{U}_{ij}(\mu_b,\mu_W) C^g_j(\mu_W)\bigr] 
\bra \pek |Q_{i}| \bar{B}^0 \ket,
\nonumber \\
&&A^{g8}_{\pek}= -\frac{G_F}{\sqrt{2}} \lambda_t  [0.727 C^{g}_{8g}(\mu_W)]  \bra \pek |Q_{8g}| \bar{B}^0 \ket,
\end{eqnarray}
where the factor 0.727 can be found in (\ref{C8RGE}) in Appendix B.
Noting that the imaginary part can appear only in the SUSY
contributions, we expand $R_{\pek}$ up to the leading order of ${\rm Re}[A^{X}]/A^{SM}$ ($X=H, \chi, g36, g8$) as follows:
\begin{eqnarray}
& &R_{\pek}=\frac{{\rm Im}[A]}{A^{SM}} \Gamma, \\
& &\Gamma \equiv
\Bigl( 1-\frac{A^H}{A^{SM}}-\frac{{\rm Re}[A^{\chi}]}{A^{SM}}-\frac{{\rm Re}[A^{g36}]}{A^{SM}}-\frac{{\rm Re}[A^{g8}]}{A^{SM}} \Bigr).
\end{eqnarray}
In our interesting parameter region, $\Gamma \simeq 1$ 
because ${\rm Re}[A^{g36}]/A^{SM}$, ${\rm Re}[A^{g8}]/A^{SM} \ll 1$, and $A^{H}/A^{SM} \simeq -{\rm Re}[A^{\chi}]/A^{SM} \simeq {\cal O}(0.1)$ numerically.
Therefore, we obtain the expression of $R_{\pek}$ up to leading order as
\begin{equation}
\Delta S_{\pek}^1 \equiv -1.46 \Bigl[ \frac{{\rm Im}[A_{\pek}^{\chi}]}{A_{\pek}^{SM}}+\frac{{\rm Im}[A_{\pek}^{g36}]}{A_{\pek}^{SM}}+\frac{{\rm Im}[A_{\pek}^{g8}]}{A_{\pek}^{SM}} \Bigr].
\end{equation}
Thus, each contribution can be considered separately.
Speaking in more detail, we have two kinds of chargino contributions.
One of them comes from the superpartner of the diagram for the charged Higgs contribution in which flavor violation is originated by $V_{CKM}$.
This almost dominates ${\rm Re}[A^{\chi}]$, and thus, we have the above relation $|{\rm Re}[A^{\chi}]| \sim |{\rm Re}[A^H]|$ in our parameter region because all the mass scales in the chargino diagram are of the same order as the mass scales in the charged Higgs diagram.
However, this gives little contribution to ${\rm Im}[A^{\chi}]$ because ${\rm Im}[(V_{CKM})_{32}] \ll {\rm Re}[(V_{CKM})_{32}]$.
The other chargino contribution is from $(m^2_{\tilde{u}_L})_{32}$, which is caused by the nonuniversality of sfermion masses.
This gives large contribution to ${\rm Im}[A^{\chi}]$.

We present our numerical results for $m_0$ dependence of $\Delta S_{\phi K_S}$ and $\Delta S_{\eta' K_S}$ in Fig.\,1.
In the numerical calculation, we use $\Delta S^X_{\phi(\eta')K}=-1.46 {\rm Im}[A^X_{\phi(\eta')K}]/A^{SM}_{\phi(\eta')K}$.
To make the result more correct, for the total $\Delta
S_{\phi(\eta')K}$ in Fig.\,1, we use Eq.\,(\ref{CS}), 
although the sum of each contribution is almost equal to the total $\Delta S_{\phi(\eta') K}$.
In the upper figures in Fig.\,1, the $C_8^{\chi}$ contribution, the
$C_8^g$ gluino contribution, and the $C_3^g$-$C_6^g$ gluino
contribution are denoted by red (deep gray), green (middle gray), and cyan (light gray) lines, respectively. 
The SUSY contributions are proportional to the $\sin\theta_{SUSY}$.
Hence, we take $\theta_{SUSY}$ to be $-\pi /2$ through our analysis so that the SUSY contributions are maximized.
In our numerical calculation, we scan the order one parameter and
phases in the matrix $V_Q$ in the range (0.8 -- 1.2) and (0 -- 2$\pi$) with fixing the $\theta_{SUSY}$.

Figure\,1 shows that if $m_0$ is much larger than $m_{30}$, the $CP$ asymmetries of $\BPK$ and $\BEK$ deviate from the SM expectation.
At the point $m_0=m_{30}=100{\rm GeV}$, 
since $CP$ violating couplings $(m^2_{\tilde{d}_L})_{32}$ and $(m^2_{\tilde{u}_L})_{32}$ vanish, $\Delta S_{\phi K_S}$ and $\Delta S_{\eta' K_S}$ become the values predicted by the SM.
$S_{\phi K_S}$ and $S_{\eta' K_S}$ have a SM contribution that originated from the Kobayashi-Maskawa phase.
From our approximation, the SM predictions of $\Delta S_{\phi K_S}$ and $\Delta S_{\eta' K_S}$ are
\begin{equation}
\Delta S_{\phi K_S}(SM)=+0.02,~~~~ \Delta S_{\eta' K_S}(SM)=+0.006.
\end{equation} 
These SM contributions are indicated by horizontal lines in the upper figures in Fig.\,1.
Figure\,1 shows that $\Delta S_{\phi K_S}$ and $\Delta S_{\eta' K_S}$ have the same sign. 
This is a consequence of absence of the RR mixing in the down-type squark sector.
As mentioned in \S~2, the chargino contribution has the same phase as the gluino one.
Numerical calculation reveals that this interference is constructive.
As we can see from Fig.\,1, the $C_{8}^g$ gluino 
contribution decreases with increasing $m_0$, while the $C_3^g$-$C_6^g$ gluino ones and $C_8^{\chi}$ chargino one do not decrease.
To understand this feature, we use the mass insertion diagrams.
The diagrams A, B, and C correspond to $C_{8}^g$ gluino contribution,
$C_3^g$-$C_6^g$ gluino contribution, and $C_8^g$ chargino contribution, respectively.
As we can find from the diagrams B and C, for $C_3^g$-$C_6^g$ gluino
and $C_8^{\chi}$ chargino contribution, $\tilde{c}_L$ and $\tilde{s}_L$ become heavy with increasing $m_0$, but this decoupling effect is cancelled by the enhancement of the flavor changing coupling $(m^2_{\tilde{d}_L(\tilde{u}_L)})_{32}$ which is proportional to ($m^2_{30}-m^2_0$).
On the other hand, since the $C_8^g$ gluino contribution has two
propagators, which include heavy sfermion ($\tilde{b}_R$ and $\tilde{s}_L$) and one flavor changing coupling, this contribution is decoupled in the limit $m_0 \to \infty$.

\vspace{5mm}
\begin{figure}[!t]
\begin{tabular}{cr}
\hspace{3mm}
\begin{minipage}{0.5\hsize}
\begin{center}
\includegraphics[width=5.5cm,clip]{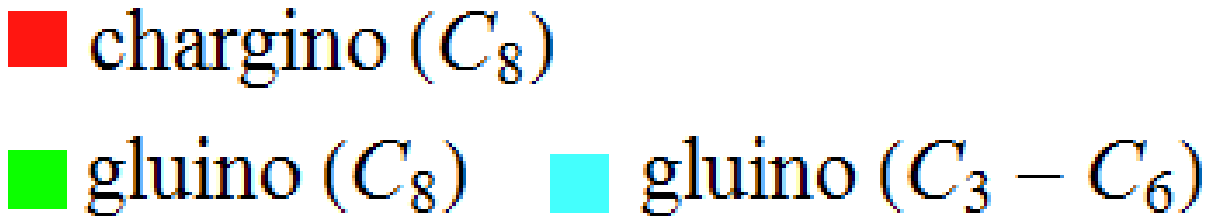}
\end{center}
\end{minipage}
\hspace{-4mm}
\begin{minipage}{0.5\hsize}
\begin{center}
\includegraphics[width=5.5cm,clip]{label.eps}
\end{center}
\end{minipage}
\end{tabular}
\vspace{-5mm}
\begin{center}
\hspace{-5mm}
\includegraphics[width=15cm,clip]{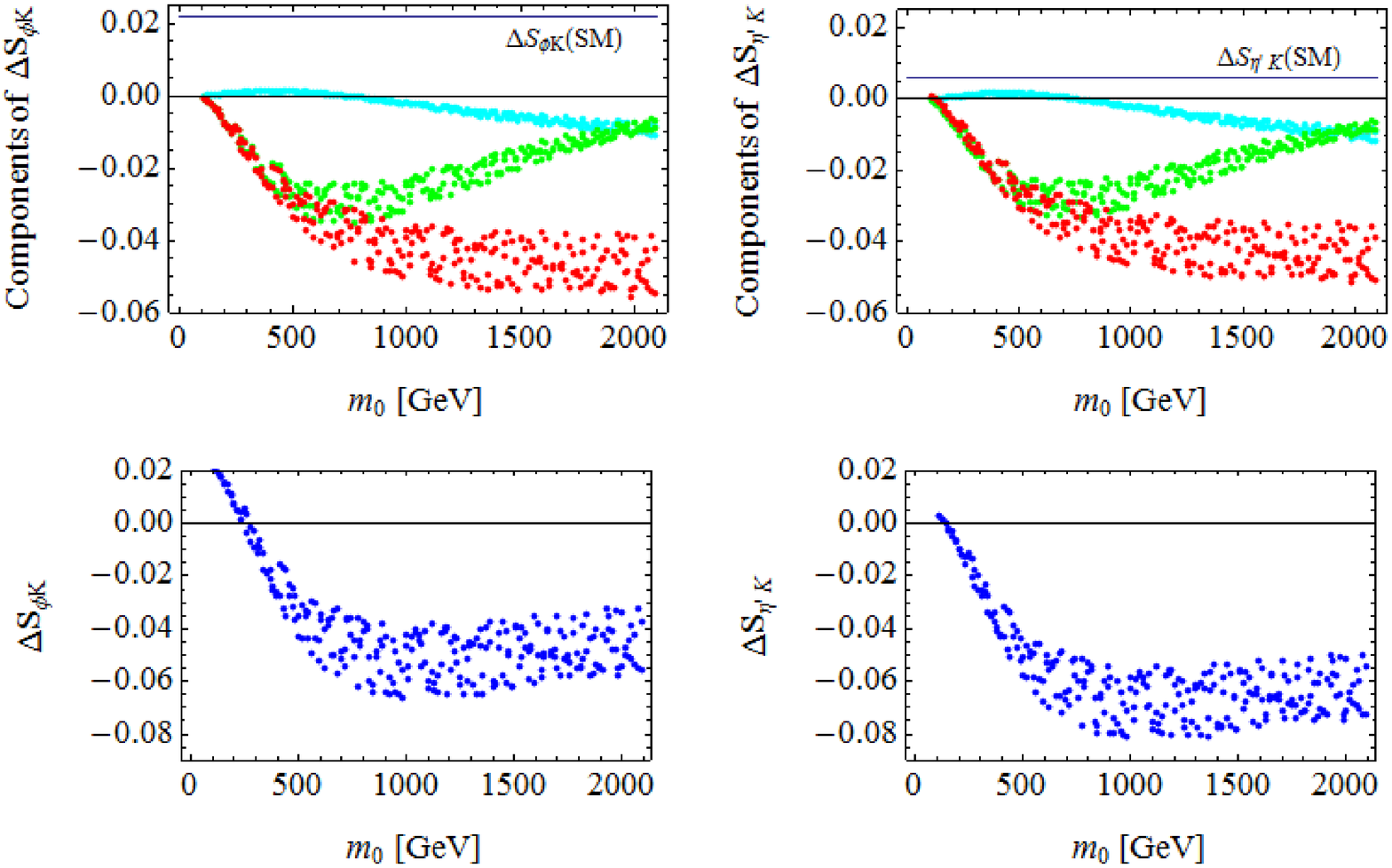}
\caption{\footnotesize $m_0$ dependence of $\Delta S_{\phi K_S}$ and $\Delta S_{\eta' K_S}$ and the SUSY contributions.
Blue, red, green, and cyan (or the order of deeper gray scale) correspond to $\Delta S_{\phi(\eta') K_S}$, $C_8^{\chi}$ chargino contribution, $C_8^g$ gluino contribution and $C_3^g$-$C_6^g$ gluino contribution, respectively.
}
\end{center}
\end{figure}

\begin{tabular}{cr}
\begin{minipage}{0.5\hsize}
\begin{flushleft}
\includegraphics[width=6.5cm,clip]{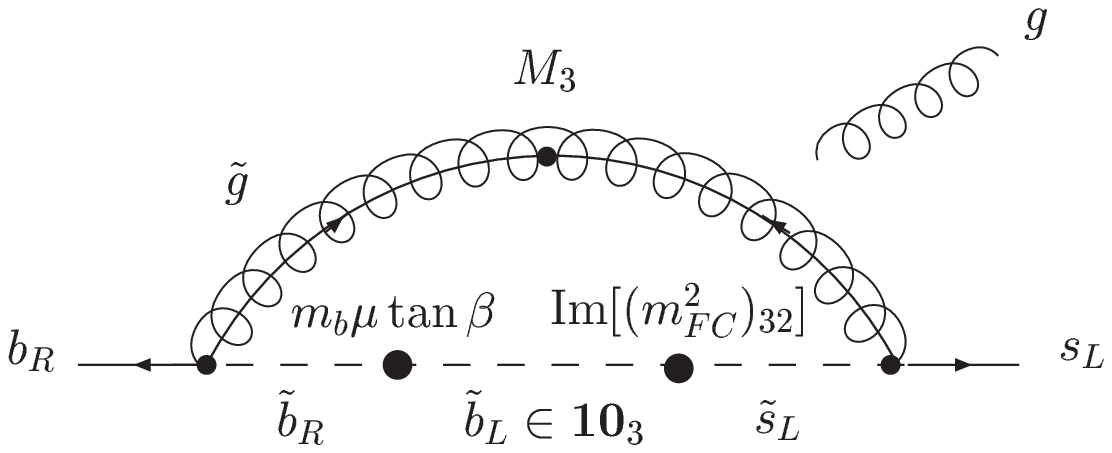}
\end{flushleft}
\end{minipage}
\hspace{-1.4cm}
\begin{minipage}{0.5\hsize}
\vspace{-3mm}
\begin{eqnarray}
&&\propto~\frac{m_b \mu \tan\beta}{m^2 m_3^2} \frac{(m^2_{30}-m_0^2) \lambda^2 \sin\theta_{SUSY}}{m^2} 
\nonumber \\
&&
\xrightarrow{m~\rightarrow~\infty}~~0\nonumber 
\end{eqnarray}
\end{minipage}
\\
Diagram~A:{\footnotesize~~~~~Contribution from ${\rm Im}[A^{g8}]$.}
\\
\\
\end{tabular}
\\
\begin{tabular}{cr}
\begin{minipage}{0.5\hsize}
\begin{flushleft}
\hspace{5mm}
\includegraphics[width=7cm,clip]{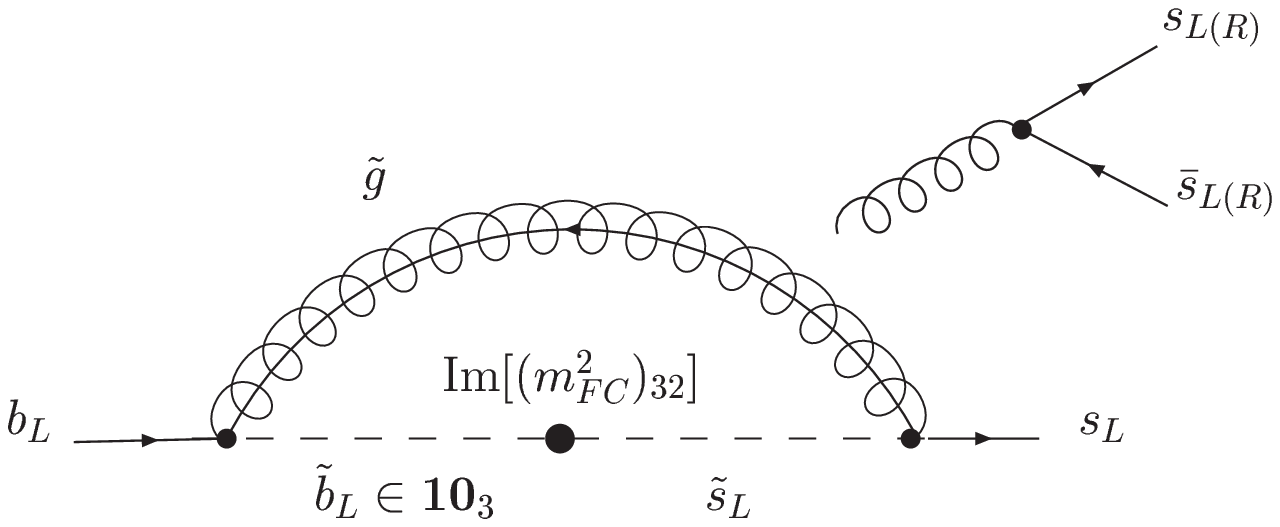}
\end{flushleft}
\end{minipage}
\hspace{-1.6cm}
\begin{minipage}{0.5\hsize}
\vspace{0mm}
\begin{eqnarray}
&&\propto~\frac{1}{m^2_3} \frac{(m^2_{30}-m_0^2) \lambda^2 \sin\theta_{SUSY}}{m^2} 
\nonumber \\
&&
\xrightarrow{m~\rightarrow~\infty}~~\frac{1}{m^2_3} \lambda^2 \sin\theta_{SUSY} \nonumber 
\end{eqnarray}
\end{minipage}
\\
\hspace{8mm}
~~~~~Diagram~B:{\footnotesize~~~~~Contribution from ${\rm Im}[A^{g36}]$.}
\\
\\
\end{tabular}
\\
\begin{tabular}{cr}
\begin{minipage}{0.5\hsize}
\hspace{5mm}
\includegraphics[width=6cm,clip]{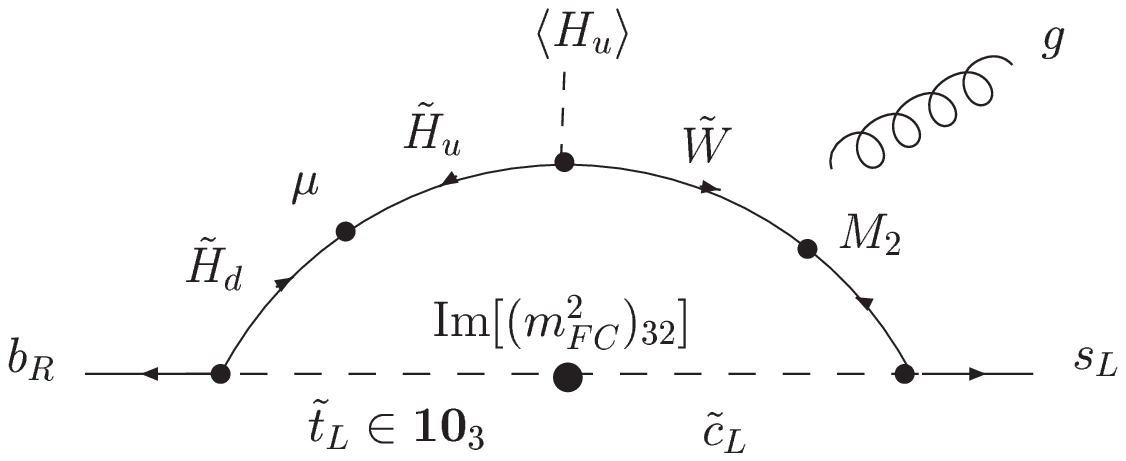}
\end{minipage}
\hspace{-1.4cm}
\begin{minipage}{0.5\hsize}
\vspace{0mm}
\begin{eqnarray}
&&\propto~\frac{m_b \tan\beta}{m^2_3 } \frac{(m^2_{30}-m_0^2) \lambda^2 \sin\theta_{SUSY}}{m^2} 
\nonumber \\
&&
\xrightarrow{m~\rightarrow~\infty}~~\frac{m_b \tan\beta}{m^2_3 } \lambda^2 \sin\theta_{SUSY} \nonumber 
\label{mix21}
\end{eqnarray}
\end{minipage}
\\
\hspace{8mm}
~~~~~Diagram~C:{\footnotesize~~~~~Contribution from ${\rm Im}[A^{\chi}]$.}
\\
\\
\end{tabular}
\\


We present the numerical results of $\mu$ dependence of $\Delta S_{\phi K_S}$ and $\Delta S_{\eta' K_S}$ in Fig.\,2.
As we can see from Fig.\,2, the $C_8^g$ gluino contribution decrease
as decreasing $\mu$, while the $C_8^{\tilde{\chi}}$ chargino contribution slightly increase.
The reason is that $C_8^g$ gluino contribution (diagram A) is
proportional to $\mu$, 
while $C_8^{\chi}$ chargino contribution (diagram C) is decoupled for large $\mu$ parameter, because higgsino states are in the loop.
The naturalness argument requires a small $\mu$ parameter; thus, the
gluino contribution cannot be large, 
so the chargino contribution and constructive interference between them is important to make the sizable deviation of $CP$ asymmetries from the SM expectation.
\begin{figure}[!h]
\begin{center}
\hspace{-5mm}
\includegraphics[width=15.5cm,clip]{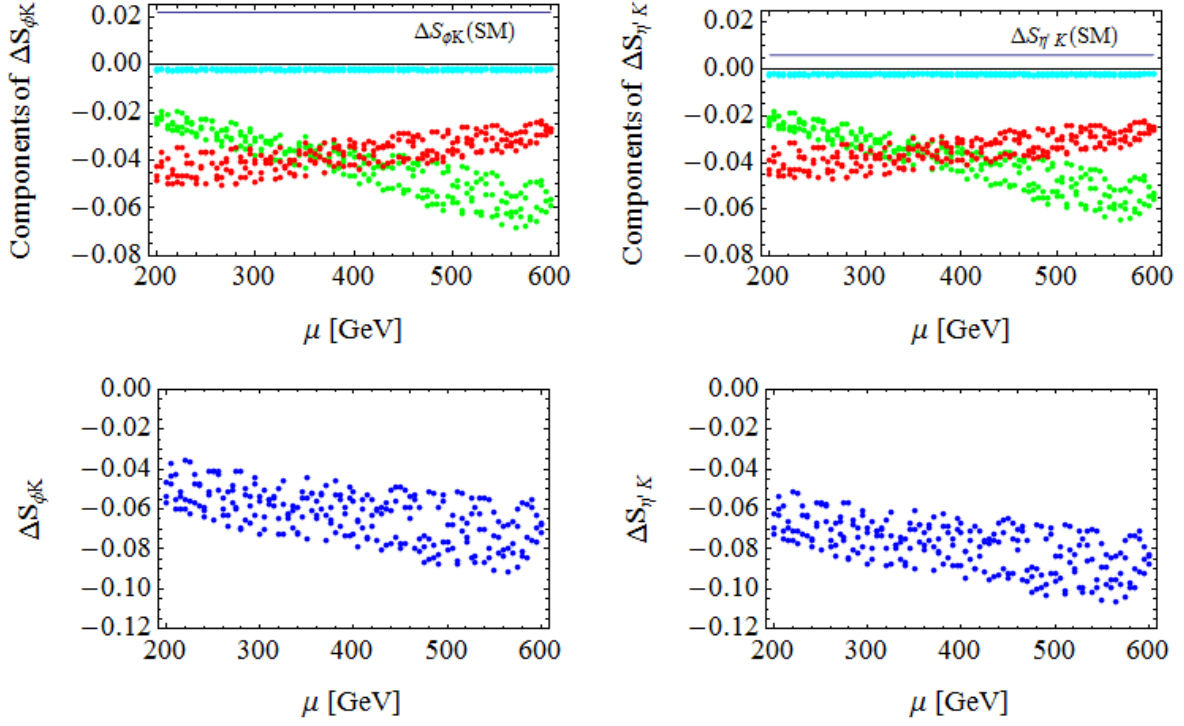}
\caption{\footnotesize $\mu$ dependence of $\Delta S_{\phi K_S}$ and $\Delta S_{\eta' K_S}$ and the SUSY contributions.}
\end{center}
\end{figure}
\begin{figure}[!h]
\begin{center}
\hspace{-5mm}
\includegraphics[width=15cm,clip]{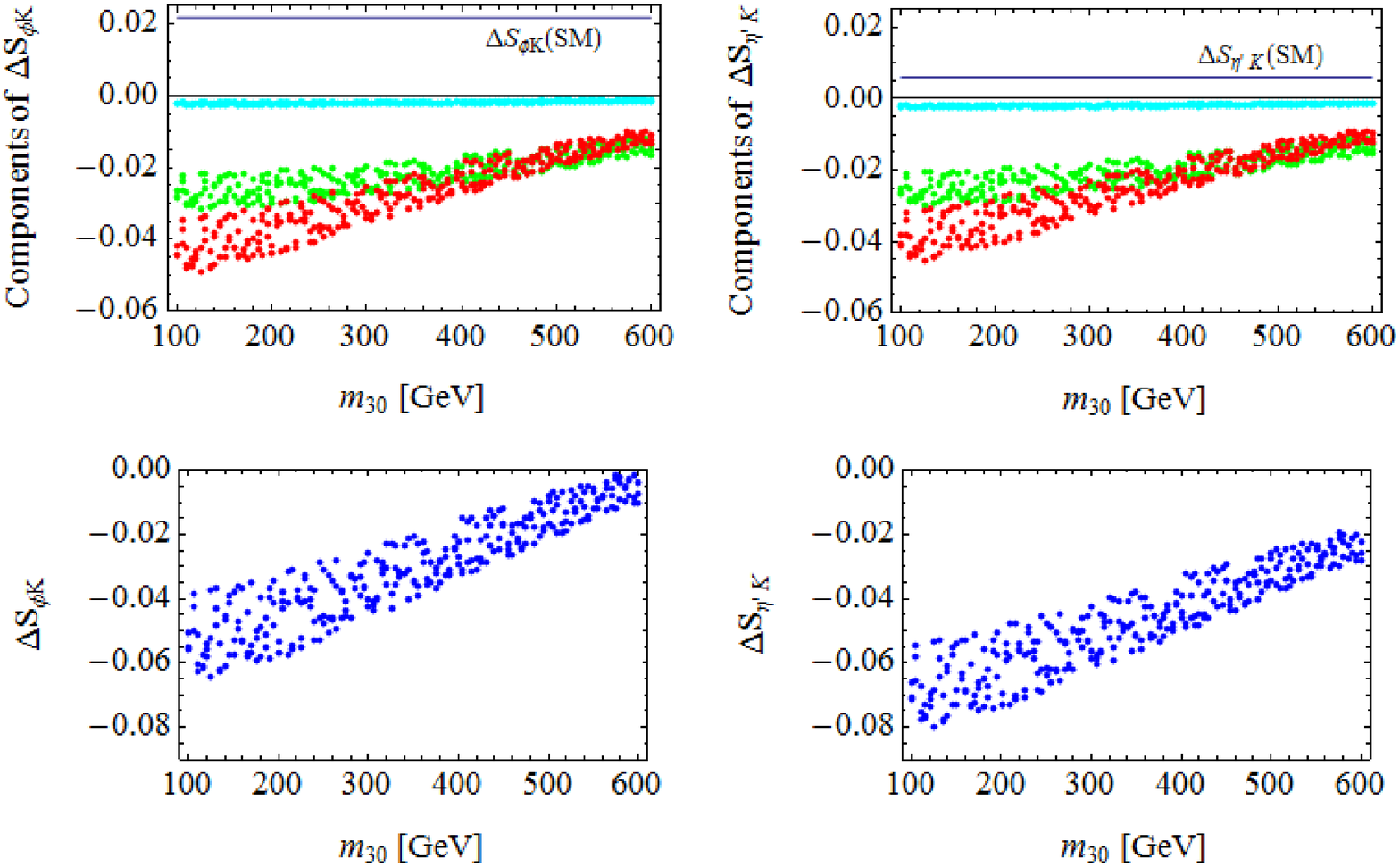}
\caption{\footnotesize  $m_{30}$ dependence of $\Delta S_{\phi K_S}$ and $\Delta S_{\eta' K_S}$ and the SUSY contributions.}
\end{center}
\end{figure}
\vspace{-5mm}
\begin{figure}[!h]
\begin{center}
\hspace{-5mm}
\includegraphics[width=14.5cm,clip]{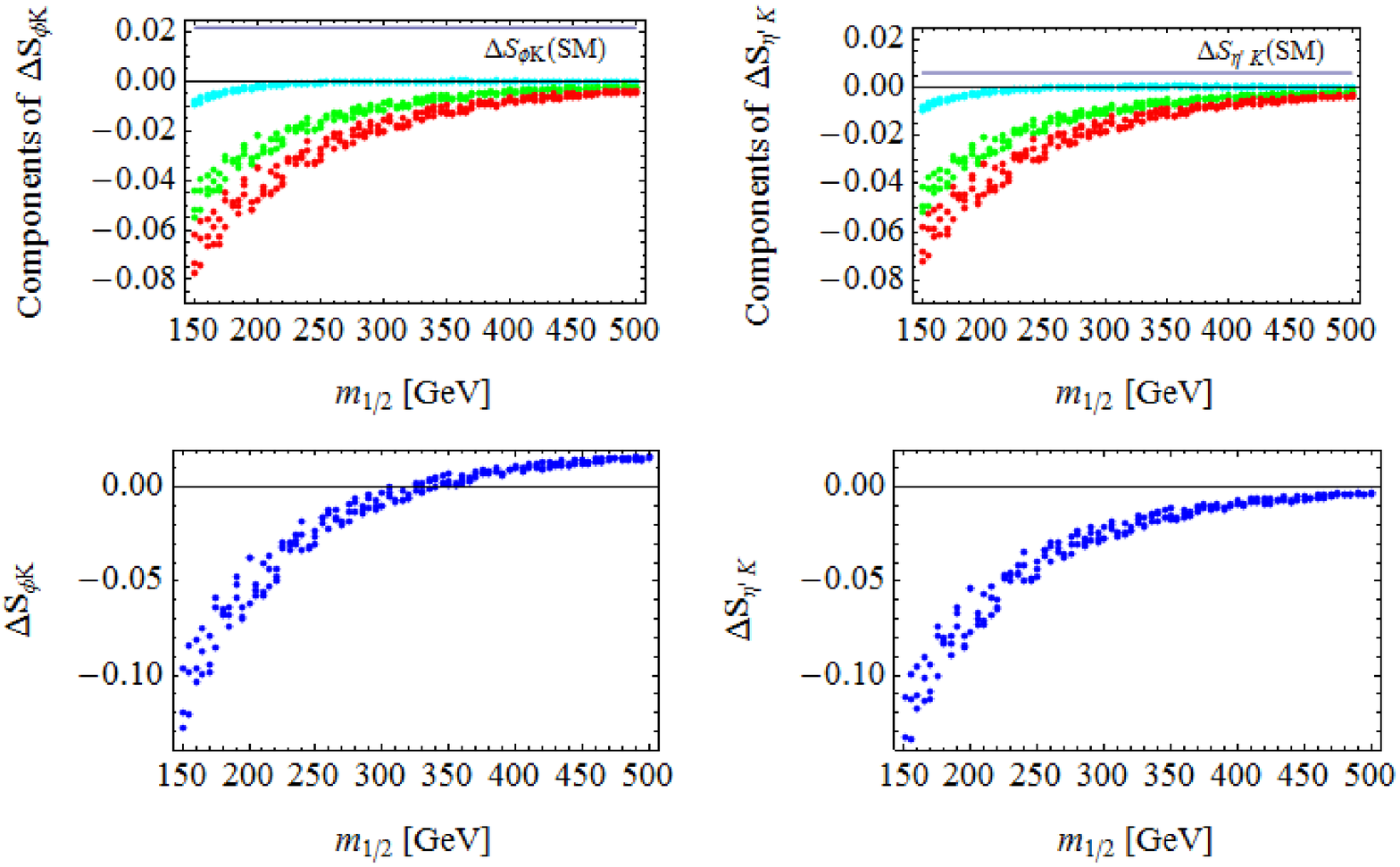}
\caption{\footnotesize  $m_{1/2}$ dependence of $\Delta S_{\phi K_S}$ and $\Delta S_{\eta' K_S}$ and the SUSY contributions.}
\end{center}
\end{figure}
\vspace{5mm}

Next, we show $m_{30}$ and $m_{1/2}$ dependences of $\Delta S_{\phi
K_S}$ and $\Delta S_{\eta' K_S}$ in Figs.\,3 and 4, respectively.
$\Delta S_{\phi K_S}$ and $\Delta S_{\eta' K_S}$ strongly depend on $m_{30}$ and $m_{1/2}$.
Again, smallness of these parameters is required by the naturalness argument.
Thus, the deviations of the $CP$ asymmetries from the SM expectation are consistent with the naturalness argument in the MSSM Higgs sector.
In Fig.\,3, we find that the $C_8^{\chi}$ chargino contribution is more enhanced than $C_8^g$ gluino one.

For the chargino contribution, since both stops ($\tilde{t}_L$ and
$\tilde{t}_R$) are included in ${\bf 10}_3$, if $m^2_3$ is around the
scale $A_t m_t$, then one of the mass eigenstates of the stops becomes much lighter than the other squarks.
As the result, $|\Delta S| \sim 0.1$ can be obtained.
Again, the chargino contribution is important in this parameter region.

Both the present experimental central values of the $CP$ asymmetries of $\BPK$ and $\BEK$ negatively deviate from the SM expectation as ${\cal O}(0.1)$.
These are consistent with our numerical results.
However, we can not extract the new physics contributions to $|\Delta
S|$ from the present experimental data because the experimental errors
are still large and it denotes that the data are almost consistent with the SM predictions.
If the errors can be reduced in the near future experiments, 
the deviations from the SM can be confirmed and we may obtain several new constrains to our model from the $CP$ asymmetries.

\section{Constraints and discussion}
In this section, we discuss the constraints from the $b \to s \gamma$ process and the electric dipole moment (EDM).
Since CP violating sources for the $\BPK$ and $\BEK$ come from the flavor changing coupling $(m^2_{\tilde{d}_L})_{32}$ and $(m^2_{\tilde{u}_L})_{32}$ in the nonuniversal mass model, we have to explore also the $b \to s \gamma$ constraint.
The experimental value of the branching ratio of $b \to s \gamma$ process $Br(b \to s \gamma)_{exp}=(3.55 \pm 0.24^{+0.09}_{-0.01} \pm 0.03) \times 10^{-4}$ \cite{HFAG} is now almost consistent with the SM prediction 
$Br(b \to s \gamma)_{SM}=(2.98 \pm 0.26) \times 10^{-4}$ for $E_\gamma > 1.6 \,{\rm GeV}$ \cite{BSGSM}, $(3.15 \pm 0.23) \times 10^{-4}$ for $E_\gamma > 1.6 \,{\rm GeV}$ \cite{BSGSM2}, $(3.57 \pm 0.49) \times 10^{-4}$ for $E_\gamma > m_b / 20$ \cite{BSGSM3}.  
Thus, we plot the region in which the total value of the coefficient $C_7^{total}(m_b)$ is consistent with the SM value $C_7^{SM}(m_b)$ at 10$\%$ level.
Figure\,5 shows the complex plane of $C_7^{total}(m_b)/C_7^{SM}(m_b)$.
We scan the $\theta_{SUSY}$ in the range (0 - 2$\pi$) in this calculation.
The grey arrow represents the contribution that includes the SM,
charged Higgs, and CKM originated chargino contributions, which cannot have the $CP$ violating phase except for a small KM phase.
As mentioned in the previous section, 
because of the cancelation between the charged Higgs and the CKM originated chargino contribution in the positive $\mu$ case,
the sum of these contributions becomes almost the same as the SM contribution as pointed out in Ref.\,\cite{LHPlightH}.
\footnote{Actually, exact cancellation is not required here, 
because there are other contributions for example, gluino and $(m^2_{\tilde{u}_L})_{32}$ originated chargino contributions.}
On the other hand, the black arrow represents the sum of the gluino 
and $(m^2_{\tilde{u}_L})_{32}$ originated chargino contributions that have a SUSY phase $\theta_{SUSY}$.
As we can see from Fig.\,5, $\theta_{SUSY} \simeq -\pi/2$ is consistent with the $b \to s \gamma$ constraint.

\begin{figure}[!t]
\begin{center}
\hspace{-5mm}
\includegraphics[width=6cm,clip]{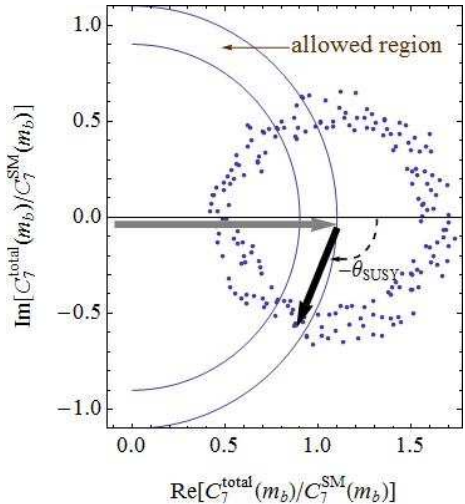}
\caption{\footnotesize $b \to s \gamma$ constraint. We use the parameter set in (\ref{pset}) and $m_{H^{\pm}}=130~{\rm GeV}$.}
\end{center}
\end{figure}

For the EDM constraints, it has been pointed out that Chromo EDMs
(CEDMs) and EDMs strongly constrain the off-diagonal entries of sfermion mass matrices in Ref.\,\cite{Hisano:2004tf}.
In the parameter region adopted in our paper, most of SUSY contributions to the (C)EDMs can be decoupled in the large $m$ limit.
However, there is a non decoupling diagram contributing to the (C)EDMs of up quark because both left and right-handed stops have small masses.   
This diagram makes a generically effective mass insertion parameter as
\begin{tabular}{cr}
\begin{minipage}{0.5\hsize}
\vspace{-25mm}
\begin{flushleft}
\vspace{3mm}
\includegraphics[width=6cm,clip]{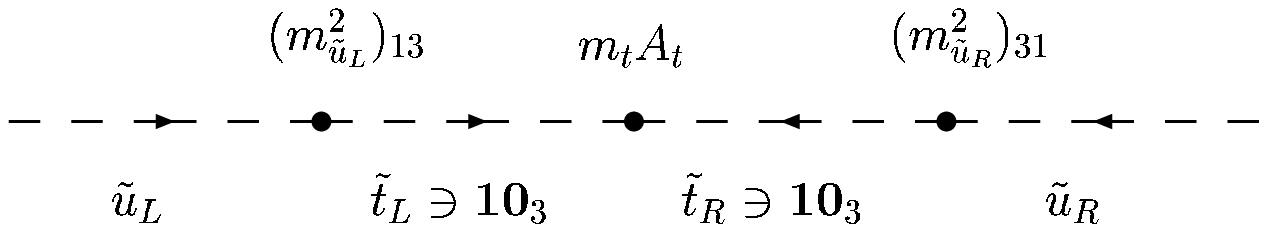}
\end{flushleft}
\end{minipage}
\hspace{-1.4cm}
\begin{minipage}{0.5\hsize}
\vspace{4mm}
\begin{eqnarray}
&\xrightarrow{~~~~~~~}& {\rm Im}[(\delta^u_{LR})_{11}] / m_3^2 \nonumber \\
&\sim& \frac{(m^2_{\tilde{u}_L})_{13}}{m^2_{\tilde{u}_L}} \frac{A_t m_t}{m^2_{\tilde{t}_L} m^2_{\tilde{t}_R}} \frac{(m^2_{\tilde{u}_R})_{31}}{m^2_{\tilde{u}_R}}  \nonumber \\
&\sim& \frac{(m_{30}^2-m_0^2)\lambda^3}{m^2} \frac{A_t m_t}{m_3^2 m_3^2} \frac{(m_{30}^2-m_0^2)\lambda^3}{m^2} \nonumber \\
&\xrightarrow{m~\rightarrow~\infty}&  \bigl(\frac{A_t m_t \lambda^6}{m_3^2} \bigr)/m_3^2~. 
\label{deltaLR}
\end{eqnarray}
\end{minipage}
\end{tabular}
\\

Although accidental cancellation in Im[$(\delta^u_{LR})_{11}$] can
satisfy these constraints, 
this issue can be solved in explicit models.
For example, we can construct such models in a framework of
spontaneous $CP$ violation in ${\rm E}_6$ GUT with horizontal symmetry \cite{E6CP}.  

Finally, we comment on the significance of our results for the specific models.
In the specific models, other tree-level contributions to the off-diagonal part of the sfermion mass matrices
exist generically. 
Especially, the correction to the 5bar squark fields can give a large effect to the result.
In this paper, however, we do not discuss them because these contributions are dependent on the explicit models
and in many cases they can be decoupled in the limit $m \gg m_3$.
If the condition $m \gg m_3$ is not satisfied  
it is possible to expect more deviation of the $CP$ asymmetry of $B \to \phi (\eta') K_S$
in such explicit models.

\section{Summary}
We investigated the $CP$ asymmetries of $\BPK$ and $\BEK$ 
in a nonuniversal mass model in which the third generation sfermion
included in ${\bf 10}$ of $SU(5)$ have different masses ($\sim m_3$) from the others ($\sim m$).
We chose MSSM parameters $m \gg m_3$, $m_{1/2}$ and $\mu$, which suppress 
the FCNCs and EDMs for the first two generation fields without destabilizing the weak scale.
Then the chargino contribution is as important as the gluino contribution.
The reasons are as follows.
For the smaller $\mu$ parameter, the gluino contribution, which is
proportional 
to $\mu\tan\beta$ becomes smaller and the chargino contribution becomes larger.
Moreover, since both stops are included in ${\bf 10}$, small $m_3$
makes one of the stops very light owing to the left-right mixing.
Owing to the $SU(2)_L$ symmetry, the chargino contribution has 
the same $CP$ phase as the gluino contribution and the numerical calculation reveal their interference is constructive.
As the result, $|\Delta S|$ becomes comparatively large.

Other $CP$ violating observables related to the $b \to s$ processes,
 for example, $CP$ violation in $b \to s \gamma$\cite{Chua:1998dx}
  and time dependent $CP$ asymmetry of $B_s \to J/\psi \phi$\cite{Dighe:1995pd},
   etc., may deviate from the SM predictions.
This is our future work.      

We found that the deviations of the CP asymmetries of $\BPK$ and $\BEK$ can be ${\cal O}(0.1)$ in this model.
We checked that both deviations $\Delta S_{\phi K}$ and $\Delta S_{\eta' K}$ have the same sign.
This is a consequence of which this model has almost only LL mixings.
These deviations may be confirmed in future experiments.

\section*{Acknowledgements}
N.M. and K.S. are supported in part by Grants-in-Aid for Scientific 
Research from the Ministry of Education, Culture, Sports, Science 
and Technology of Japan. The work of T.Y. is supported by the 21st Century COE
Program of Nagoya University.

\newpage

\appendix
%
\section{Hadronic Matrix elements}

Here, we give the expression of hadronic matrix elements for $\BPK$ and $\BEK$ in the method of the naive factorization.
For $\BPK$, the matrix elements with the operator $Q_i$ given in (\ref{Q}) are given as \cite{Khalil,Ali}
\begin{eqnarray}
&&\bra \phi \bar{K}^0 | Q_1 | \bar{B}^0 \ket = 0,~~~~~
\bra \phi \bar{K}^0 | Q_2 | \bar{B}^0 \ket = 0,~~~~~
\bra \phi \bar{K}^0 | Q_3 | \bar{B}^0 \ket = \frac{4}{3}X,  \nonumber \\
&&\bra \phi \bar{K}^0 | Q_4 | \bar{B}^0 \ket = \frac{4}{3}X,~~~~~
\bra \phi \bar{K}^0 | Q_5 | \bar{B}^0 \ket = X,~~~~~
\bra \phi \bar{K}^0 | Q_6 | \bar{B}^0 \ket = \frac{1}{3}X, 
\end{eqnarray}
and
\begin{eqnarray}
&&\bra \phi \bar{K}^0 | Q_{7\gamma} | \bar{B}^0 \ket = 0, \nonumber \\
&&\bra \phi \bar{K}^0 | Q_{8g} | \bar{B}^0 \ket = -\frac{\alpha_s}{4 \pi} \frac{m_b}{\sqrt{\bra q^2 \ket}}
\bigl[ \bra Q_4 \ket +\bra Q_6 \ket -\frac{1}{3}(\bra Q_3 \ket+\bra Q_5 \ket)   \bigr],
\label{QgBPK}
\end{eqnarray}
where $X$ is defined as
\begin{equation}
X = 2 F_+^{B \to K} (m^2_{\phi}) f_{\phi} m_{\phi} (p_K \cdot \epsilon_{\phi}).
\end{equation}
Here, $m_{\phi}$ represents the $\phi$ meson mass, $f_{\phi}$ is decay constant of the $\phi$ meson, $F_+^{B \to K}$ is the transition form factor which is estimated at the $m_{\phi}$ scale, $p_K$ is momentum of the K meson, and $\epsilon_{\phi}$ represents  the polarization vector of the $\phi$ meson.
We used $m_{\phi} = 1.02 $~GeV, $f_{\phi}=0.233$~GeV, $F_+^{B \to K}=0.35$, and 
\begin{equation}
(p_K \cdot \epsilon_{\phi}) = \frac{m_B}{m_{\phi}} 
\sqrt{\bigl[ \frac{1}{2 m_B} (m_B^2 -m_K^2 +m_{\phi}^2)  \bigr]^2-m^2_{\phi}}~,
\end{equation}
where $m_B$ and $m_K$ are $B$ meson mass and $K$ meson mass, respectively.
$\bra q^2 \ket$ in (\ref{QgBPK}) is the average of momentum carried by virtual gluon in $Q_{8g}$.
Kinematical consideration leads to the physical range $m_b^2/4 \le  \bra q^2 \ket \le m_b^2/2$.
We use $\bra q^2 \ket=m_b^2/4$ numerically, in this case SUSY contributions to $CP$ asymmetries are maximized.
Hadronic matrix elements for the operator $\tilde{Q}_i$ are given by
\begin{eqnarray}
\bra \phi \bar{K}^0 | \tilde{Q}_i | \bar{B}^0 \ket = \bra \phi \bar{K}^0 | Q_i | \bar{B}^0 \ket. 
\label{tildeQBPK}
\end{eqnarray}
This relation is derived from the parity invariance of the strong
interaction and the fact 
that the initial and final states have the same parity.

Hadronic matrix elements for $\BEK$ are given by \cite{Khalil,Ali}
\begin{eqnarray}
&&\bra \eta' \bar{K}^0 | Q_1 | \bar{B}^0 \ket = X_2,~~~~~
\bra \eta' \bar{K}^0 | Q_2 | \bar{B}^0 \ket = \frac{1}{3} X_2, \nonumber \\
&&\bra \eta' \bar{K}^0 | Q_3 | \bar{B}^0 \ket = \frac{1}{3} X_1 + 2X_2 +\frac{3}{4}X_3,~~~
\bra \eta' \bar{K}^0 | Q_4 | \bar{B}^0 \ket = X_1 + \frac{2}{3}X_2 +\frac{3}{4}X_3,  \nonumber \\
&&\bra \eta' \bar{K}^0 | Q_5 | \bar{B}^0 \ket = \frac{R_1}{3}X_1 - 2X_2 -\bigl( 1-\frac{R_2}{3} \bigr)X_3,  \nonumber \\
&&\bra \eta' \bar{K}^0 | Q_6 | \bar{B}^0 \ket = R_1 X_1 -\frac{2}{3} X_2 -\bigl( \frac{1}{3} - R_2 \bigr)X_3,  
\end{eqnarray}
and
\begin{eqnarray}
&&\bra \eta' \bar{K}^0 | Q_{7\gamma} | \bar{B}^0 \ket = 0, \nonumber \\
&&\bra \eta' \bar{K}^0 | Q_{8g} | \bar{B}^0 \ket = -\frac{\alpha_s}{4 \pi} \frac{m_b}{\sqrt{\bra q^2 \ket}}
\bigl[ \bra Q_4 \ket +\bra Q_6 \ket -\frac{1}{3}(\bra Q_3 \ket+\bra Q_5 \ket)   \bigr],
\end{eqnarray}
where 
\begin{eqnarray}
&&X_1 = -(m_B^2 -m_{\eta'}^2) F_1^{B \to \pi} (m^2_K) \frac{X_{\eta'}}{\sqrt{2}} f_K , \nonumber \\
&&X_2 = -(m_B^2 -m_{\eta'}^2) F_1^{B \to K} (m^2_{\eta'}) \frac{X_{\eta'}}{\sqrt{2}} f_K , \nonumber \\
&&X_3 = -(m_B^2 -m_{\eta'}^2) F_1^{B \to K} (m^2_{\eta'}) \sqrt{2 f_K^2 - f_{\pi}^2}Y_{\eta'}, \nonumber \\
&&R_1=\frac{2 m_K^2}{(m_b-m_s)(m_s+m_d)},~~~~~R_2=\frac{2m_K^2-m_{\pi}^2}{(m_b-m_s)m_s},
\end{eqnarray}
where $F_1^{B \to \pi}(q^2)$ is $B-\pi$ transition form factor which
is estimated at the  $q^2$ scale, $f_K$,\,and $f_{\pi}$ are decay constants of $K$ and $\pi$ meson, respectively. 
$X_{\eta'}$, $Y_{\eta'}$ represent the rate of the $u\bar{u}+d\bar{d}$ and $s\bar{s}$ component in the $\eta'$ meson, respectively.
We use $F_1^{B \to \pi}(m_K^2)=0.35$, $f_K=0.16$~GeV, $f_{\pi}=0.13$~GeV, $X_{\eta'}=0.57$ and $Y_{\eta'}=0.82$.
Since $\eta'$ contains a small $u\bar{u}$ component, $\bra \eta' \bar{K}^0 | Q_{1(2)} | \bar{B}^0 \ket$ has a non zero value.
Thus, the $\BEK$ process has a small contribution from a color suppressed tree diagram. 
Matrix elements for the operator $\tilde{Q}_i$ are given by
\begin{equation}
\bra \eta' \bar{K}^0 | \tilde{Q}_i | \bar{B}^0 \ket = -\bra \eta' \bar{K}^0 | Q_i | \bar{B}^0 \ket,
\label{tildeQBEK}
\end{equation}
where the minus sign is a consequence of the parity difference of the initial and final states.

By using the above formulae, numerical values of the matrix elements are found as follows: 

\vspace{3mm}
\begin{tabular}{|l|c|c|c|c|c|c|c|} 
\hline
      & $Q_1$ & $Q_2$ &$Q_3$ &$Q_4$ &$Q_5$ &$Q_6$ &$Q_{8g}$ \\ \hline
$\BPK$ & 0     & 0     & 2.89 & 2.89 & 2.17 & 0.72 &$-$0.068 \\ \hline
$\BEK$ &$-$0.506 &$-$0.169 &$-$3.17 &$-$2.90 & 1.55 & $-$1.96 &0.135 \\ 
\hline
\end{tabular} 
\vspace{1cm}

\section{Wilson coefficients}

In this Appendix, we give the expressions of the high energy ($\mu_W
\simeq {\cal O}(m_W)$) Wilson coefficients and of the evolution matrices 
that are used to calculate the low energy ($\mu_b \simeq {\cal O}(m_b)$) Wilson coefficients from the high energy ones.

Weak scale Wilson coefficients $C_i(\mu_W)$ are constructed from the
SM contributions $C_i^{SM}(\mu_W)$, charged Higgs contributions
$C_i^{H}(\mu_W)$, gluino contributions $C_i^{g}(\mu_W)$, and chargino contributions $C_i^{\chi}(\mu_W)$ as 
\begin{eqnarray}
&&C_1=C_1^{SM},~~~~~~~~~~~~~~~~~~~~~~~~~~~~~C_2=C_2^{SM}, \nonumber \\
&&C_3=C_3^{SM}+C_3^{g},~~~~~~~~~~~~~~~~~~~~~~C_4=C_4^{SM}+C_4^{g}, \nonumber \\
&&C_5=C_5^{SM}+C_5^{g},~~~~~~~~~~~~~~~~~~~~~~C_6=C_6^{SM}+C_6^{g}, \nonumber \\
&&C_{7\gamma} = C_{7\gamma}^{SM}+C_{7\gamma}^{H}+C_{7\gamma}^{g}+C_{7\gamma}^{\chi},~~~
C_{8g} = C_{8g}^{SM}+C_{8g}^{H}+C_{8g}^{g}+C_{8g}^{\chi}.
\end{eqnarray}
Here, we consider only the SM contributions for the non FCNC process $C_{1(2)}$.
Moreover, the chargino contributions are taken into account for only $C_{7\gamma}$ and $C_{8g}$ in which the SM contributions
are suppressed by chirality suppression.
${\cal O}(\mu_W)$ scale Wilson coefficients are given as follows. \cite{Bertolini,Wang,Murayama}
\vspace{3mm}
\\
{\bf Standard Model contributions:}
\begin{eqnarray}
&&C_1^{SM} = \frac{14 \alpha_s}{16 \pi},~~~~~~~~~~C_2^{SM}=1-\frac{11 \alpha_s}{24 \pi},~~~~~~~~C_3^{SM}=-\frac{\alpha_s}{24 \pi} E(x_t),
\nonumber \\
&&C_4^{SM}=\frac{\alpha_s}{8 \pi} E(x_t),~~~~~C_5^{SM}=-\frac{\alpha_s}{24 \pi} E(x_t),~~~~~C_6^{SM}=\frac{\alpha_s}{8 \pi} E(x_t),
\nonumber \\
&&C_{7\gamma}^{SM}=-x_t \bigl( F_1(x_t)+\frac{3}{2}F_2(x_t) \bigr),~~~~~~~~~~~C_{8g}^{SM}=-\frac{3}{2}x_t F_1(x_t),
\end{eqnarray}
where $x_t=m_t^2/m_W^2$,
\begin{eqnarray}
&&E(x)=\frac{x(x^2+11x-18)}{12(x-1)^3}+\frac{-9x^2+16x-4}{6(x-1)^4}\log x, \nonumber \\
&&F_1(x) = \frac{x^3-6x^2+3x+2+6x \log x}{12(x-1)^4},
\nonumber \\ &&
F_2(x) = \frac{2 x^3+3 x^2 -6x+1 -6x^2 \log x}{12(x-1)^4}.
\end{eqnarray}
\vspace{3mm}
\\
{\bf Charged Higgs contributions:}
\begin{eqnarray}
&&C_{7\gamma}^{H}=-\frac{x_H}{2} \bigl(\bigl( \frac{2}{3} F_1(x_H)+F_2(x_H) \bigr)\cot^2\beta +\frac{2}{3}F_3(x_H)+F_4(x_H) \bigr),
\nonumber \\
&&C_{8g}^{H}=-\frac{x_H}{2} \bigl( F_1(x_H)\cot^2\beta + F_3(x_H) \bigr),
\end{eqnarray}
where $x_H=m_t^2/m_H^2$, $m_H$ is the charged Higgs mass,
\begin{equation}
F_3(x) = \frac{x^2-4x+3+2x \log x}{2(x-1)^3}, 
~~~~~
F_4(x) = \frac{x^2 -1 -2x \log x}{2(x-1)^3}.
\end{equation}
\vspace{3mm}
\\
{\bf Gluino contributions:}
\begin{eqnarray}
C_3^{g}&=&\frac{-\sqrt{2}}{4 G_F \lambda_t} \frac{\alpha_s^2}{M_3^2}
\Bigl( \sum_{k,l} \Gamma^{*ks}_{DL} \Gamma^{kb}_{DL} \Gamma^{*ls}_{DL} \Gamma^{ls}_{DL} 
\Bigl[ -\frac{1}{9} B_1(x_{kg},x_{lg}) -\frac{5}{9} B_2(x_{kg},x_{lg}) \Bigl] 
\nonumber \\ &&
+ \sum_{k} \Gamma^{*ks}_{DL} \Gamma^{kb}_{DL} \Bigl[ -\frac{1}{18} C_1(x_{kg}) +\frac{1}{2} C_2(x_{kg}) \Bigl] \Bigr),
\nonumber \\
C_4^{g}&=&\frac{-\sqrt{2}}{4 G_F \lambda_t} \frac{\alpha_s^2}{M_3^2}
\Bigl( \sum_{k,l} \Gamma^{*ks}_{DL} \Gamma^{kb}_{DL} \Gamma^{*ls}_{DL} \Gamma^{ls}_{DL} 
\Bigl[ -\frac{7}{3} B_1(x_{kg},x_{lg}) +\frac{1}{3} B_2(x_{kg},x_{lg}) \Bigl] 
\nonumber \\ &&
+ \sum_{k} \Gamma^{*ks}_{DL} \Gamma^{kb}_{DL} \Bigl[\frac{1}{6} C_1(x_{kg}) -\frac{3}{2} C_2(x_{kg}) \Bigl] \Bigr),
\nonumber \\
C_5^{g}&=&\frac{-\sqrt{2}}{4 G_F \lambda_t} \frac{\alpha_s^2}{M_3^2}
\Bigl( \sum_{k,l} \Gamma^{*ks}_{DL} \Gamma^{kb}_{DL} \Gamma^{*ls}_{DR} \Gamma^{ls}_{DR} 
\Bigl[ \frac{10}{9} B_1(x_{kg},x_{lg}) +\frac{1}{18} B_2(x_{kg},x_{lg}) \Bigl] 
\nonumber \\ &&
+ \sum_{k} \Gamma^{*ks}_{DL} \Gamma^{kb}_{DL} \Bigl[ -\frac{1}{18} C_1(x_{kg}) +\frac{1}{2} C_2(x_{kg}) \Bigl] \Bigr),
\nonumber \\
C_6^{g}&=&\frac{-\sqrt{2}}{4 G_F \lambda_t} \frac{\alpha_s^2}{M_3^2}
\Bigl( \sum_{k,l} \Gamma^{*ks}_{DL} \Gamma^{kb}_{DL} \Gamma^{*ls}_{DR} \Gamma^{ls}_{DR} 
\Bigl[ -\frac{2}{3} B_1(x_{kg},x_{lg}) +\frac{7}{6} B_2(x_{kg},x_{lg}) \Bigl] 
\nonumber \\ &&
+ \sum_{k} \Gamma^{*ks}_{DL} \Gamma^{kb}_{DL} \Bigl[\frac{1}{6} C_1(x_{kg}) -\frac{3}{2} C_2(x_k,{kg}) \Bigl] \Bigr),
\nonumber \\
C_{7\gamma}^{g}&=&\frac{-\sqrt{2}}{4 G_F \lambda_t} \frac{\alpha_s \pi}{M_3^2}  \sum_{k}
\Big(\Gamma^{*ks}_{DL} \Gamma^{kb}_{DL}\Bigl[-\frac{4}{9}D_1(x_{kg}) \Bigr] 
- \frac{M_3}{m_b}  \Gamma^{*ks}_{DL} \Gamma^{kb}_{DR} \Bigl[ -\frac{4}{9}D_2(x_{kg}) \Bigr]  \Big)
\nonumber \\
&=&\frac{\sqrt{2} }{2 G_F \lambda_t} (-\frac{4}{9}) \sum_{k} \frac{ \alpha_s \pi}{m^2_{\tilde{d}_{k}}}
\Bigl( \Gamma^{kb}_{DL} \Gamma^{*ks}_{DL} F_2(x_{gk}) - \frac{M_3}{m_b}\Gamma^{kb}_{DR} \Gamma^{*ks}_{DL} F_4(x_{gk}) \Bigr),
\nonumber \\
C_{8g}^{g}&=&\frac{-\sqrt{2}}{4 G_F \lambda_t} \frac{\alpha_s \pi}{M_3^2}  \sum_{k}
\Big(\Gamma^{*ks}_{DL} \Gamma^{kb}_{DL}\Bigl[-\frac{1}{6}D_1(x_{kg})+\frac{3}{2}D_3(x_{kg}) \Bigr] 
\nonumber \\ &&
- \frac{M_3}{m_b}  \Gamma^{*ks}_{DL} \Gamma^{kb}_{DR} \Bigl[ -\frac{1}{6}D_2(x_{kg})+\frac{3}{2}D_4(x_{kg}) \Bigr]  \Big)
\nonumber \\
&=&\frac{\sqrt{2} }{2 G_F \lambda_t} \sum_{k} \frac{ \alpha_s \pi}{m^2_{\tilde{d}_{k}}}
\Bigl( \Gamma^{kb}_{DL} \Gamma^{*ks}_{DL} \Bigl[ -\frac{3}{2} F_1(x_{gk}) -\frac{1}{6} F_2(x_{gk})  \Bigr] 
\nonumber \\
&&- \frac{M_3}{m_b}\Gamma^{kb}_{DR} \Gamma^{*ks}_{DL} \Bigl[ -\frac{3}{2} F_3(x_{gk}) -\frac{1}{6} F_4(x_{gk}) \Bigl] \Bigr),
\end{eqnarray}
where $x_{kg}$=$x^{-1}_{gk}=m^2_{\tilde{d}_k} / M_3^2$, $M_3$ is the gluino mass, $m^2_{\tilde{d}_k}$ represents a mass eigen value of the down-type squarks.
We define a unitary matrix of the down-type squark mass matrix $\Gamma_D=(\Gamma_{DL}, \Gamma_{DR})$ as 
\begin{equation}
(\Gamma_D \tilde{m}^2_{\tilde{d}} \Gamma_D^{\dagger})_{kl} = m^2_{\tilde{d}_k} \delta_{kl},
~~~~~~~~~~
\tilde{d}_{L(R)} = \Gamma^{\dagger}_{DL(R)} \tilde{d}_k,
\end{equation}
where $\Gamma_{DL(R)}$ is $6 \times 3$ matrix and 
$\tilde{m}^2_{\tilde{d}}$ is the scalar down mass matrix at the weak scale, which can be written as
\begin{equation}
\tilde{m}^2_{\tilde{d}}=
\begin{pmatrix}
m_{\tilde{d}_L}^2 - (\frac{1}{2} - \frac{1}{3}\sin^2\theta_W) m_Z^2 \cos2\beta + m_d^2 & -m_d (A_d^* + \mu \tan\beta) \\
-m_d (A_d + \mu^* \tan\beta) & m_{\tilde{d}_R}^2 -  \frac{1}{3}\sin^2\theta_W m_Z^2 \cos2\beta + m_d^2
\end{pmatrix}.
\end{equation}
The functions appeared above are given by
\begin{eqnarray}
B_1(x,y)&=&\int^{\infty }_0 \frac{-\frac{1}{4} z^2 dz}{(z+1)^2(z+x)(z+y)} 
\nonumber \\ 
&=&-\frac{x^2 \log x}{4(x-y)(x-1)^2}-\frac{y^2 \log y}{4(y-x)(y-1)^2}-\frac{1}{4(x-1)(y-1)},
\nonumber \\
B_2(x,y)&=&\int^{\infty }_0 \frac{z dz}{(z+1)^2(z+x)(z+y)} 
\nonumber \\ 
&=&-\frac{x \log x}{(x-y)(x-1)^2}-\frac{y \log y}{(y-x)(y-1)^2}-\frac{1}{(x-1)(y-1)},
\nonumber \\ 
C_1(x)&=&\frac{2 x^3-9 x^2+18x-11-6\log x}{36(1-x)^4},
\nonumber \\ 
C_2(x)&=&\frac{-16 x^3+45 x^2-36x +7 +6x^2(2x-3)\log x}{36(1-x)^4},
\nonumber \\ 
D_1(x)&=&\frac{-x^3+6x^2-3x-2-6x\log x}{6(1-x)^4},
\nonumber \\ 
D_2(x)&=&\frac{-x^2+1+2x\log x}{(x-1)^3},
\nonumber \\ 
D_3(x)&=&\frac{2x^3+3x^2-6x+1-6x^2\log x}{6(1-x)^4},
\nonumber \\ 
D_4(x)&=&\frac{-3x^2+4x-1+2x^2\log x}{(x-1)^3}.
\end{eqnarray}
\vspace{3mm}
\\
{\bf Chargino contributions:}
\begin{eqnarray}
C_{7\gamma}^{\chi}&=& \frac{\sqrt{2} \alpha_W \pi }{ 2 G_F \lambda_t} \sum_j \sum_{k} \frac{1}{m^2_{\tilde{u}_{k}}}
\bigl((G_{UL}^{jkb} -H_{UR}^{jkb})(G_{UL}^{*jks} -H_{UR}^{*jks}) [F_1(x_{\tilde{\chi}^-_j \tilde{u}_k})+\frac{2}{3}F_2(x_{\tilde{\chi}^-_j \tilde{u}_k})] 
\nonumber \\
&& - H_{UL}^{jkb} (G_{UL}^{*jks} -H_{UR}^{*jks}) \frac{m_{\tilde{\chi}^-_j}}{m_b}[F_3(x_{\tilde{\chi}^-_j \tilde{u}_k})+\frac{2}{3}F_4(x_{\tilde{\chi}^-_j \tilde{u}_k})]  \bigr),
\nonumber \\
C_{8g}^{\chi}&=& \frac{\sqrt{2} \alpha_W \pi }{ 2 G_F \lambda_t} \sum_j \sum_{k} \frac{1}{m^2_{\tilde{u}_{k}}}
\bigl((G_{UL}^{jkb} -H_{UR}^{jkb})(G_{UL}^{*jks} -H_{UR}^{*jks}) F_2(x_{\tilde{\chi}^-_j \tilde{u}_k})
\nonumber \\
&& - H_{UL}^{jkb} (G_{UL}^{*jks} -H_{UR}^{*jks}) \frac{m_{\tilde{\chi}^-_j}}{m_b} F_4(x_{\tilde{\chi}^-_j \tilde{u}_k})  \bigr),
\end{eqnarray}
where $x_{\tilde{\chi}^-_j \tilde{u}_k} = m^2_{\tilde{\chi}^-_j}/m^2_{\tilde{u}_k}$,
 and $\alpha_W$ is defined from $SU(2)_L$ gauge coupling, $g$, as $\alpha_W = g^2/(4 \pi)$. 
$m_{\tilde{\chi}^-_j}$ and $m_{\tilde{u}_k}$ are mass eigen values of charginos and scalar up quarks, respectively.
$G_{UL}$, $H_{UL}$,and $H_{UR}$ are defined as
\begin{eqnarray}
&&G_{UL}^{jki}=V_{j1}^* \Gamma_{UL}^{ki},~~~~~~~H_{UL}^{jki}=U_{j2}(\Gamma_{UL} \hat{Y}_d)^{ki}/g,
\nonumber \\
&&H_{UR}^{jki}=V_{j2}^*(\Gamma_{UR} \hat{Y}_u V_{CKM})^{ki}/g,
\end{eqnarray} 
where $ \Gamma_U =(\Gamma_{UL}, \Gamma_{UR})$ is a unitary matrix of the scalar up mass matrix 
which is defined in the same manner as the scalar down mass matrix,  
and $V$ and $U$ are diagonalizing matrices of the chargino mass matrix which are defined as
\begin{eqnarray}
(U m_C V^{\dagger})_{ij} = m_{\tilde{\chi}_{i}} \delta_{ij},~~~~~~
m_C=
\begin{pmatrix}
M_2 & \sqrt{2} m_W \sin\beta \\
\sqrt{2} m_W \cos\beta & \mu
\end{pmatrix}.
\end{eqnarray}

Next, we have to estimate the Wilson coefficients at the scale ${\cal O}(\mu_b)$ from the Wilson coefficients at the scale ${\cal O}(\mu_W)$
using the renormalization group equations \cite{Buras}.
To this end, we can use the evolution matrix as follows. 
\begin{equation}
C_i(\mu_b) = \sum_{j=1}^6 \hat{U}_{ij}(\mu_b,\mu_W) C_j(\mu_W).~~~~(i=1,\dots 6)~
\end{equation}
Here, we take account of only the leading order QCD corrections for simplicity.
At the leading order, the evolution matrix $\hat{U}(\mu_b,\mu_W)$ is estimated as
\begin{equation}
\hat{U}(\mu_b,\mu_W)=\hat{V} \Bigl( \Big[ \frac{\alpha_s(\mu_W)}{\alpha_s(\mu_b)} \Bigr]^{\frac{\gamma_D^{(0)}}{2 \beta_0}} \Bigr)_D \hat{V}^{^-1},
\end{equation}
where $\beta_0=\frac{11}{3}N_C -\frac{2}{3}f$ is the coefficient of the one loop $\beta$ function of the QCD gauge coupling.
Here, $N_C$ and $f$ are number of color and number of active flavor at the scale $\mu$ ($\mu_b<\mu<\mu_W$), respectively.
$\gamma_D^{(0)}=\hat{V}^{-1} \gamma^{(0)T} \hat{V}$ is the diagonalized matrix of $\gamma^{(0)}$ which is given by
\begin{equation}
\gamma^{(0)}=\begin{pmatrix}
-2&6&0&0&0&0\\
6&-2&{-2 \over 9}&{2 \over 3}&{-2 \over 9}&{2 \over 3}\\
0&0&{-22 \over 9}&{22 \over 3}&{-4 \over 9}&{4 \over 3}\\
0&0&6-{2f \over 9}&-2+{2f \over 3}&{-2f \over 9}&{2f \over 3}\\
0&0&0&0&2&-6\\
0&0&{-2f \over 9}&{2f \over 3}&{-2f \over 9}&-16+{2f \over 3}
\end{pmatrix}.
\end{equation}
Using $N_C=3$, $f=5$, $\alpha(\mu_b)=0.22$, $\alpha(\mu_W)=0.1176$ \cite{Amsler:2008zz}
 we obtain 
\begin{equation}
\hat{U}(\mu_b,\mu_W)=\begin{pmatrix}
1.117&-0.267&0&0&0&0\\
-0.267&1.117&0&0&0&0\\
-0.0014&0.0120&1.134&-0.210&0.0079&0.0791\\
0.0034&-0.0273&-0.305&0.987&-0.0197&0.0541\\
-0.0010&0.0079&0.0107&0.0376&0.928&0.0541\\
0.0039&-0.0341&-0.0484&-0.162&0.309&1.692\\
\end{pmatrix}.
\end{equation}
For $C_{7\gamma}(m_b)$ and $C_{8g}(m_b)$, we can use the following formulae \cite{Buras}
\begin{eqnarray}
&&C_{7\gamma}(m_b) = 0.695C_{7\gamma}(\mu_W) + 0.085 C_{8g}(\mu_W) -0.158 C_2(\mu_W), \nonumber \\
&&C_{8g}(m_b) = 0.727 C_{8g}(\mu_W) -0.074 C_2 (\mu_W) .
\label{C8RGE}
\end{eqnarray}

\end{document}